\documentclass[aps,amsmath,amssymb,prb,10pt,twocolumn,showpacs,letterpaper,superscriptaddress]{revtex4-1}
\usepackage{graphicx,txfonts,color,tcolorbox,ulem}
\usepackage{bm}
\usepackage{mathrsfs}
\usepackage{comment}
\usepackage{mathtools}
\usepackage{hyperref}
\usepackage{epstopdf}
\usepackage{letltxmacro}
\usepackage{appendix}

\newcommand{\g}{\gamma}
\newcommand{\de}{\delta}

\newcommand{\ka}{\kappa}

\newcommand{\p}{\pi}

\newcommand{\f}{\phi}
\newcommand{\F}{\Phi}

\newcommand{\w}{\omega}
\newcommand{\W}{\Omega}
\newcommand{\De}{\Delta}
\newcommand{\G}{\Gamma}
\renewcommand{\S}{\Sigma}

\newcommand{\pd}{\partial}

\newcommand{\round}[1]{\left( #1 \right)}

\newcommand{\curly}[1]{\left\{#1\right\}}
\newcommand{\abs}[1]{\left| #1 \right|}

\newcommand{\beq}{\begin{equation}}
\newcommand{\eeq}{\end{equation}}
\newcommand{\Beq}{\begin{eqnarray}}
\newcommand{\Eeq}{\end{eqnarray}}
\newcommand{\bml}{\begin{multline}}

\newcommand{\bea}{\begin{align}}
\newcommand{\ena}{\end{align}}
\newcommand{\bsp}{\begin{split}}
\newcommand{\esp}{\end{split}}

\newcommand{\bS}{{\boldsymbol{S}}}

\newcommand{\ez}{\hat{\boldsymbol z}}

\renewcommand{\bm}{{\boldsymbol m}}

\DeclareMathOperator{\sgn}{sgn}

\newcommand{\tI}{\tilde{I}}

\newcommand{\vf}{\varphi}

\begin{document}
\title{Detecting spin current noise in quantum magnets with photons}
\author{Joshua Aftergood}
\affiliation{Department of Physics, Queens College of the City University of New York, Queens, NY 11367, USA}
\affiliation{Physics Doctoral Program, The Graduate Center of the City University of New York, New York, NY 10016, USA}
\author{Mircea Trif}
\affiliation{Institute for Interdisciplinary Information Sciences, Tsinghua University, Beijing 100084, China}
\author{So Takei}
\affiliation{Department of Physics, Queens College of the City University of New York, Queens, NY 11367, USA}
\affiliation{Physics Doctoral Program, The Graduate Center of the City University of New York, New York, NY 10016, USA}
\date{\today}

\begin{abstract}
A minimally invasive technique is proposed for detecting spin current noise across a junction between two quantum magnets using a high-quality microwave resonator coupled to a transmission line which is impedance matched to a photon detector downstream. Photons in the microwave resonator couple inductively to the spins in the spin subsystem, and the noise in the junction spin current imprints itself into the output photons propagating along the transmission line. The technique is capable of extracting both the dc and finite frequency noise via the output photon flux and also opens doors to the studies of photon counting statistics and to the possible generation of non-classical radiation produced by spin current fluctuations at a quantum magnet junction.
\end{abstract}
\maketitle


\section{Introduction}

Shot noise in mesoscopic conductors arises as a consequence of the quantized nature of charge transport~\cite{blanterPRP00}. Its utility far exceeds that of the equilibrium (Johnson-Nyquist) counterpart, as shot noise can be used to extract the charge~\cite{heiblumPSS06} and the statistics of relevant charge carriers as well as to probe quantum many-body effects~\cite{selaPRL06} and entanglement~\cite{beenakkerPRL03}. In insulators, charge fluctuations are gapped out but localized electron spins can still generate fluctuations in pure spin current, or spin current noise. It is then anticipated that these pure spin currents can reveal nontrivial properties of the underlying spin system in analogy with the above-mentioned charge scenario. Indeed, recent theoretical inquiries on pure spin current noise in insulating magnets have purported its utility in revealing the quantum uncertainty associated with magnon eigenstates~\cite{kamraPRL16}, the non-trivial spin scattering and heating processes generated at a detector interface~\cite{matsuoPRL18} and the effective spin and statistics of the tunneling spin quasiparticles~\cite{aftergoodPRB18}. 

Insulating materials present novel concerns. In metallic systems, spin current noise detection is possible via its observable effects on simultaneous charge current fluctuations~\cite{mishchenkoPRB03sn,belzigPRB04,meairPRB11,arakawaPRL15}; similar methods are clearly not available for insulators. The prevailing spin current detection method in the latter has been the inverse spin Hall effect wherein spin current is detected electrically by coupling a metal with strong spin-orbit interactions to the active spin system~\cite{sinovaRMP15}. However, this spin Hall detection scheme, while reliable for spin current detection, may be unreliable for the measurement of spin current noise because spin Hall conversion processes can result in noise enhancement~\cite{dragomirovaEPL08}. Other techniques exist for detecting spin fluctuations including spin noise spectroscopy~\cite{sinitsynRPP16}, SQUID-based spectroscopy~\cite{granataPR16}, and quantum-impurity relaxometry~\cite{vandersarNATC15,casolaNRM18,flebusPRL18}. However, spin noise spectroscopy does not measure the current-current correlator, i.e., spin current noise, and neither SQUID-based approaches nor relaxometry are amenable to extracting current-current correlations in the spin sector either. Therefore, in connecting theory to experiment, a measurement technique capable of directly detecting spin current noise is of unique interest.

\begin{figure}[t]
\includegraphics[width=\linewidth]{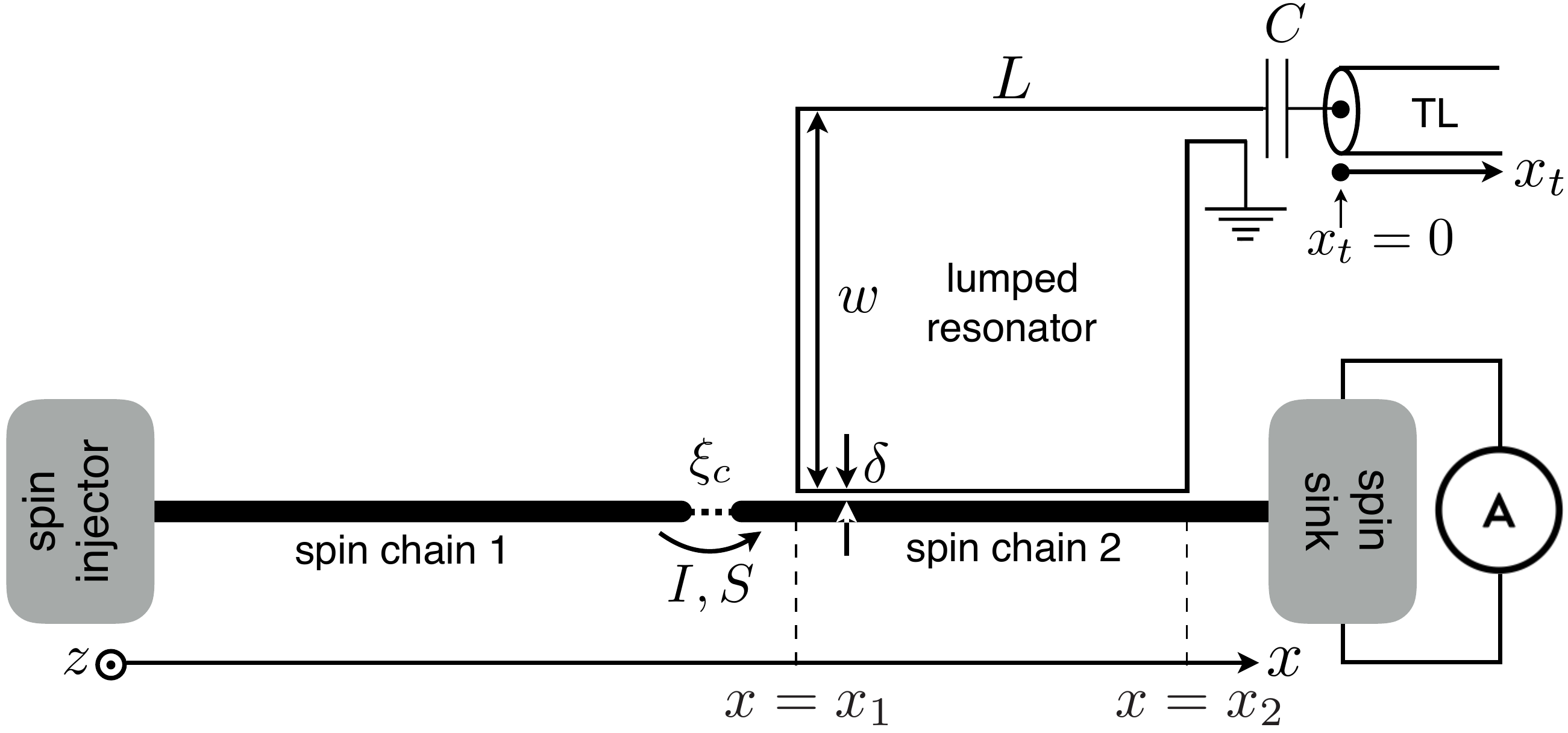}
\caption{A depiction of the proposed system. A microwave resonator, comprised of a loop of inductance $L$ and a capacitor of capacitance $C$, is set in the $xy$ plane a distance $\de$ from spin chain $2$, which is coupled with strength $\xi_c$ to spin chain $1$.}
\label{fig1}
\end{figure}

In this work, we show that a microwave resonator circuit can be used to directly measure nonequilibrium pure spin current noise generated in a quantum magnet. Here, we illustrate this possibility in the context of a one-dimensional quantum antiferromagnet chain. As shown in Fig.~\ref{fig1}, we consider a situation in which one spin chain (chain 1) is driven out of equilibrium by spin injection at its open end with the downstream end weakly exchange-coupled to a second spin chain (chain 2) set near a microwave cavity. The microwave cavity couples inductively to chain 2, and measurements are transmitted electrically into a transmission line where the photon number flux encodes spin current fluctuations across the coupled spin chain subsystem. We show that by coupling spins to light it is possible to measure the junction spin current noise without resort to the inverse spin Hall effect. The proposed setup also opens doors to the studies of photon counting statistics and the possible generation of non-classical radiation produced by spin current fluctuations at a quantum magnet junction.

\section{Heuristic picture}

We first qualitatively illustrate the mode of operation. Let us consider spin current tunneling between two coupled quantum antiferromagnet chains with uniaxial spin symmetry along the $z$ axis (see Fig.~\ref{fig1}). Spin injection, facilitated by, e.g., the spin Hall effect at the upstream end of chain 1, establishes a nonequilibrium spin bias between the two chains, and the resulting spin current flowing across the chains can be absorbed and measured at the right metal reservoir via the inverse spin Hall effect.

\begin{figure}[t]
\includegraphics[width=0.7\linewidth]{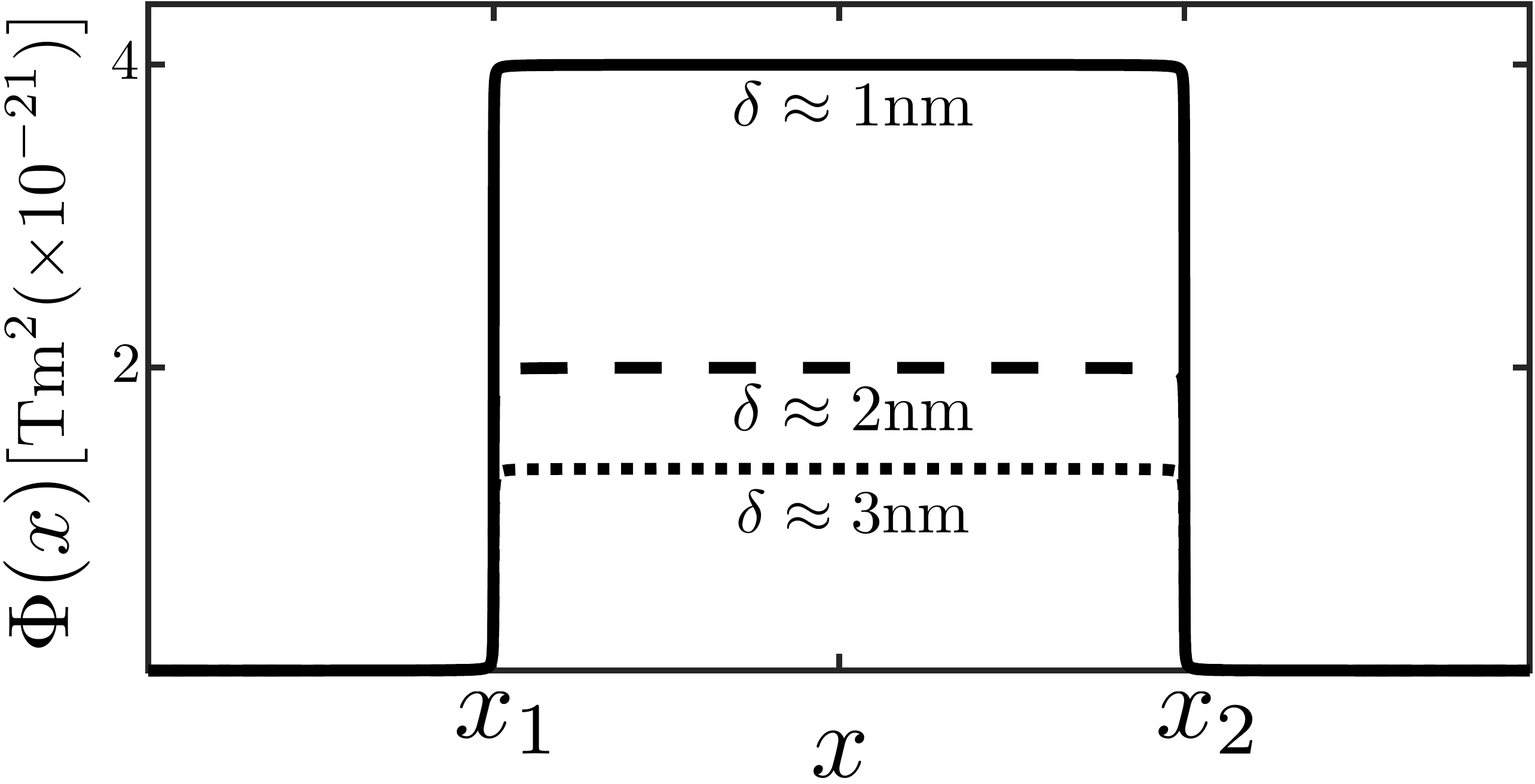}
\caption{A plot depicting magnetic flux through the loop as spins enter spin chain $2$. The sharp drop-offs indicate that spins outside of the ``influence region" $x_1<x<x_2$ are effectively ignored by the microwave resonator.}
\label{fig1b}
\end{figure}

A rectangular wire loop with inductance $L$ is placed a distance $\de$ away from chain 2, and is oriented so that it lies in the $xy$ plane and its bottom edge stretches from $x=x_1$ to $x=x_2$. In this geometry, the magnetic flux through the wire loop sharply increases by one unit when a single $z$-polarized spin-1 quasiparticle enters the ``influence region" defined by $x_1<x<x_2$ on chain 2. The magnitude of that unit depends on the distance $\de$ and the width $w$ of the loop. Considering the spin-1 quasiparticle as a magnetic dipole $\bm=-\g\hbar\ez$ located at position $x$ on chain 2 ($\g$ being the gyromagnetic ratio), the flux through the loop reads
\beq
\F(x)= \frac{-\mu_0m_z}{4 \pi} \int_{0}^{s}\int_0^w[(x'-x)^2 + (y+\de)^2]^{-3/2}dx'dy,
\eeq
where $s=x_2 - x_1$ is the influence region. Fig.~\ref{fig1b} shows a sharp increase in the flux as the quasiparticle tunnels into the influence region for various $\de$. We find that the flux is essentially independent of the quasiparticle position in the influence region and that the dipolar fields from spins located outside of the influence region are effectively irrelevant to the total flux through the loop.

Imagine now that the $x=x_1$ edge of the wire loop is located at site $i_1$ on chain 2 and the $x=x_2$ edge at site $i_2$. Then the term in the Hamiltonian modeling the tunneling of spin-1 quasiparticles from site $i_1$ to $i_1+1$ on chain 2 (allowed by the intrachain exchange interaction) involves a term of the form $S^-_{2,i_1}S^+_{2,i_1+1}$ ($S^+_{2,i_1}S^-_{2,i_1+1}$), where $S^\pm_{\nu,i}$ denotes the usual spin raising (lowering) operator on chain $\nu$ at site $i$. However, the fact that every tunneling process is accompanied by a flux change $\F$ in the loop requires that the tunneling operator is modified to $S^+_{2,i_1}S^-_{2,i_1+1}e^{i\F q_0/\hbar}+h.c.$, where $q_0$ is an operator obeying $[q_0,\f_0]=i\hbar$ and translates the influence flux $\f_0$ through the loop by $\F$, i.e., 
\beq
e^{\pm i\F q_0/\hbar} \f_0 e^{\mp i\F q_0/\hbar}=\f_0\mp\F.
\eeq
In this configuration, $q_0$ is the charge on the capacitor $C$ (see Fig.~\ref{fig1}), and $\f_0$ and $q_0$ form a conjugate variable pair. The spin tunneling operator endowed by the flux translation operator indicates that the spin tunneling process involves interactions with the electromagnetic environment formed by the loop-capacitor subsystem. In a similar fashion, the spin tunneling term at the $x=x_2$ edge is also modified to $S^+_{2,i_2}S^-_{2,i_2+1}e^{-i\F q_0/\hbar}+h.c.\ $.


Energetic considerations show that the wire loop in principle affects spin transport in chain 2. That is, for every unit of flux $\F$ tunneling into the influence region, the energy of the inductive system increases by $E_\F=\F^2/2L$, where $L$ is the inductance of the resonator. As a result, two regimes emerge: what we call the {\it noninvasive} and {\it invasive} regimes. In the {\it noninvasive} regime, $E_\Phi$ is much smaller than the non-equilibrium bias $\mu$, the temperature $T$, and the photon frequency $\Omega$, i.e., $E_\Phi\ll \mu,k_BT,\hbar\Omega$ (note that we will define the photon frequency later). In this regime, we may focus exclusively on the effect of nonequilibrium spin transport on the electromagnetic environment and neglect the back-action of the environment on the spin subsystem. A fluctuating spin current at the junction leads to a fluctuating magnetic flux through the wire loop and thus to a fluctuating electromotive force inside the resonator by Faraday's law of induction. As mentioned previously, this fluctuating electrical signal is ultimately detected in the transmission line via an output photon number flux containing a direct imprint of the spin current noise. 

In the {\it invasive} regime, the environmental effect is not negligible and the spin transport in chain 2 should deviate from their unperturbed values. For $E_\F$ much greater than the nonequilibrium spin bias $\mu$ and temperature $T$, i.e., $\mu,k_BT\ll E_\F$, tunneling events become increasingly unfavorable energetically due to the resistive electromotive force emerging from Lenz's law and leads to a suppression in the spin transmission through the chain. We refer to this phenomenon as {\it inductive blockade}, which may be thought of as the magnetic analog of the well-known Coulomb blockade studied extensively in quantum conductors~\cite{devoretPRL90,ingoldCHA92}. If the junction resistance is strong enough to suppress elastic scattering between the nodes, tunneling quasiparticles must have sufficient energy to excite environmental modes and proceed inelastically. However, the regime may be challenging to realize in practice due to the weakness of the spin-light interaction. In the remainder of the work, we focus on the noninvasive regime and present the technical calculations to establish the above heuristic results.

\section{Microscopic theory}

We consider two identical semi-infinite $xxz$ quantum antiferromagnet chains coupled together at their finite ends, one additionally coupled inductively to a microwave resonator and the resonator itself placed in series with a transmission line on which measurements are performed. The spin chains are modeled by the usual $xxz$ Hamiltonian 
\begin{multline}
\label{h0}
H_\nu=J\sum_{j=0}^\infty\Big[\tfrac{1}{2}S^+_{\nu,j}S^-_{\nu,j+1}\round{\de_{j,i_1}e^{i\F q_0/\hbar}+\de_{j,i_2}e^{-i\F q_0/\hbar}}\de_{\nu,2}\\
+h.c.+\Delta S_{\nu,j}^zS_{\nu,j+1}^z\Big]\ , 
\end{multline}
where $\bS_{\nu,j}$ is the spin-1/2 operator on chain $\nu$ at site $j$, $J$ is the intra-chain exchange scale and $\de_{i,j}$ is the kronecker delta. The exponential factors encode the coupling of spin chain 2 to the electromagnetic environment, as discussed previously. We assume throughout that the spin chains are in the so-called gapless phase, i.e., $|\De|<1$, where it can be suitably described using the Luttinger liquid formalism~\cite{giamarchiBOOK04}.

We model the transmission line as an infinite array of parallel LC resonators with lineic capacitance $c$, lineic inductance $l$, and characteristic impedance $z = \sqrt{l/c}$. Its Hamiltonian reads
\beq
H_{TL}=\int_0^\infty dx_t\ \curly{\frac{\f^2(x_t)}{2l}+\frac{[\pd_{x_t}q(x_t)]^2}{2c}}\ , 
\eeq
where $x_t$ labels the position along the transmission line, $q(x_t)$ and $\f(x_t)$ denote the local charge and flux, respectively, and $[q(x_t),\f(x'_t)]=i\hbar\de(x_t-x'_t)$. Located at the end of the transmission line, i.e., at $x_t=0$ (see Fig.~\ref{fig1}), is the lumped series LC resonator with capacitance $C$ and inductance $L$, and governed by the Hamiltonian
\beq
H_r= \frac{\f_0^2}{2L}+\frac{q_0^2}{2C},
\eeq
where $q_0\equiv q(x_t=0)$ and $\f_0\equiv\f(x_t=0)$. The total Hamiltonian for the full system then reads $H=\sum_\nu H_\nu+H_r+H_{TL}+V(t)$, where
\beq
V(t)=-\tfrac{1}{2}J_c(S^+_{1,0}S^-_{2,0}e^{i\mu t/\hbar}+h.c.)
\eeq
describes the tunneling of spin-1 quasiparticles across the spin chains allowed by the weak interchain exchange interaction $J_c$. The oscillatory factor $e^{i\mu t/\hbar}$ captures the fact that the spin chemical potential in chain 1 has been raised to $\mu>0$ via spin injection at its upstream end.


The standard input-output approach~\cite{clerkRMP10} proceeds by first expanding the local charge $q(x_t,t)$ in Fourier series
\begin{multline}
\label{q}
q(x_t,t)=\int_0^\infty\frac{d\w}{2 \pi}\sqrt{\frac{\hbar}{2\w z}}\ \Big[a_o(\w)e^{i\w(x_t/v-t)}\\
+a_i(\w)e^{-i\w(x_t/v+t)}+h.c.\Big]\ ,
\end{multline}
where $v = (lc)^{-1/2}$ and $a_{i,o}$ are the incoming and outgoing photon fields on the transmission line. 
Noting that the Heisenberg equations of motion evaluated at the lumped resonator $x_t=0$ give 
\begin{align}
\f_0(t) &=L\dot q_0(t) \\
 \dot\f_0 &= \frac{\pd_x q(x_t\rightarrow0,t)}{c} -\frac{q_0(t)}{C} - \frac{\Phi}{\hbar} [\tI^+_{i_1}(t)-\tI^-_{i_2}(t)],
 \end{align}
one can solve for the output photon field in terms of the known input photons,
\beq
\label{io}
a_o(\w)=-\frac{\G^*(\w)}{\G(\w)}a_i(\w)+\frac{\F}{\hbar L}\sqrt{\frac{2z\w}{\hbar}}\frac{\tI^+_{i_1}(\w)-\tI^-_{i_2}(\w)}{\G(\w)}\ .
\eeq
Here, 
\beq
\tI^\pm_i(t)=i(J/2)S^+_iS^-_{i+1}e^{\pm i\Phi q_0/\hbar}+h.c.
\eeq
is the operator for the bulk spin current in chain 2 flowing to the right at site $i$ (the exponential factor encoding the effect of the electromagnetic environment), $\G(\w)=\w^2-\W^2+i\ka\w$, $\ka = z/L$ is the rate at which photons decay into the transmission line and $\W = (LC)^{-1/2}$ is the resonance frequency. 

The first term in Eq.~\eqref{io} then corresponds to the reflection of incoming photons while the second term describes the emission or absorption of additional photons caused by the tunneling of spin-1 quasiparticles into and out of the influence region. The incoming photons are assumed to be equilibrated at resonator temperature $T_t$, which may be distinct from temperature $T$ of the spin chains, and obey
\beq
\langle a_i^\dag(\w)a_i(\w')\rangle_0=2\pi n_B(\w)\de(\w-\w')\ ,
\eeq
where $n_B(\w)=(e^{\hbar\w/k_BT_t}-1)^{-1}$ is the Bose-Einstein distribution describing the thermal photons in the transmission line.

As discussed previously, noninvasive detection of the spin current noise is conducted in the limit where spin-photon coupling strength quantified by $\Phi$ remains sufficiently small so as to leave the spin chain subsystem approximately unperturbed while the coupling of the resonator to chain 2 is simultaneously kept strong enough for detection. Solving for the output photon flux using Eq.~\eqref{io} is difficult, in principle, because the photon field itself enters the spin current operator $\tI^\pm_i(t)$ through $q_0(t)$ and this calls for self-consistency. However, since the correction to the output photon flux arising from the spin subsystem is anticipated to be suppressed by an overall multiplicative factor proportional to $\Phi^2$, c.f. second term in Eq.~\eqref{io}, the noninvasive assumption allows us to set the spin-photon coupling to zero, i.e., $\Phi=0$, when computing spin transport quantities, thus effectively lifting the self-consistency requirement.

With this in mind, let us now examine the output photon flux spectrum $\langle a^\dag_o(\w)a_o(\w')\rangle$ by inserting Eq.~\eqref{io} into the expectation value while invoking the noninvasive approximation mentioned above. The output photon flux spectrum then reads
\begin{multline}
\label{bbo}
\langle a^\dag_o(\w)a_o(\w)\rangle=n_B(\w,T_t)+\frac{2E_\F\ka\hbar\w}{\hbar^4|\G(\w)|^2}\\
\times[(1+n_B(\w,T_t))(S_{i_1}(\w,\mu,T)+S_{i_2}(\w,\mu,T))\\
-n_B(\w,T_t)(S_{i_1}(-\w,\mu,T)+S_{i_2}(-\w,\mu,T))]\ ,
\end{multline}
where $S_i(\w,\mu,T)=\int dte^{-i\w t}\left<I_i(t)I_i(0)\right>$ denotes the spin current noise at site $i$ on chain 2; we remind the reader that the transmission line (detector) temperature $T_t$ is distinguished from the spin temperature $T$. Here, the local spin current operator $I_i(t)\equiv\tI^\pm_i(t,\Phi=0)$ is now defined in the absence of the electromagnetic environment and the expectation values are taken by setting $\Phi=0$ in $H_2$ [see Eq.~\eqref{h0}]. In obtaining Eq.~\eqref{bbo}, we also ignored cross-correlations between spin current fluctuations at $x_1$ and $x_2$. This should be a good approximation because nonlocal spin correlations are expected to decay exponentially with distance at finite temperature~\footnote{Note that for $T=4~{\rm K}$ Eq.~\eqref{cor} allows for an estimate of the length scale of non-local correlations, and we find that correlations are suppressed on the order of $\abs{x_2 - x_1} > 10~{\rm nm}$.}.

In the gapless phase $|\De|<1$, each semi-infinite spin chain can be described as a chiral Luttinger liquid~\cite{aftergoodPRB18,giamarchiBOOK04}
\beq
H_\nu= \frac{\hbar u}{4\p K} \int_{-\infty}^\infty dx\ [\pd_x\vf_{\nu}(x)]^2\ ,
\eeq
where the chiral boson field $\vf_\nu(x)$ obeys 
\begin{align}
[\vf_{\nu}(x),\vf_{\nu}(x')] &=i\p K\sgn(x-x')\ ,
\end{align}
the boson speed and the Luttinger parameter, respectively~\cite{johnsonPRA73}, are given by
\begin{align}
u &= \frac{\p Ja\sqrt{1-\De^2}}{2\hbar\cos^{-1}\De} \\
K &=\frac{1}{2-(2/\p)\cos^{-1}\De}\ ,
\end{align}
and $a$ is the lattice constant. In this case, spin current inside chain 2 is essentially carried by (ballistic) noninteracting bosonic modes so the spin current fluctuations produced at the junction remain unmodified as they propagate downstream to sites $i_1$ and $i_2$~\cite{aftergoodPRB18,crepieuxPRB03}. Therefore, the local spin current noise at any point in the bulk of chain 2, i.e., $S_i(\w,\mu)$, is given by the background Johnson-Nyquist noise $S_0(T)$ plus the noise generated at the tunnel junction between the two spin chains $dS(\w,\mu,T)$. The total spin current noise at any site $i$ is then given by $S_i(\w,\mu,T)=S_0(T)+dS(\w,\mu,T)$, where~\cite{aftergoodPRB18}
\beq
\label{dS}
dS(\w,\mu,T)=2\xi_c^2\int_{-\infty}^\infty dt\cos\round{\frac{\mu t}{\hbar}}D(t,T)\,e^{-i\w t}\ ,
\eeq
with
\beq
\label{cor}
D(t,T)=\curly{\frac{\frac{\p k_BT\eta}{u\hbar}}{\sin\left[\frac{\p k_BT}{u\hbar}(iut+\eta)\right]}}^{2/K},
\eeq
$\xi_c=J_ca/2\p\eta$, $\eta \sim k_F^{-1}$ is the short distance cutoff of the theory, and $k_F$ is the Fermi wavevector~\cite{sirkerJMP12}.

\section{Noise detection}

Equation~\eqref{bbo} shows that a single mode electromagnetic environment can imprint transport quantities generated at a junction coupling two quantum magnets on the number of total output photons propagating along an attached transmission line. Reference~\onlinecite{inomataNATC16} provides an example of a single microwave photon detector based on a superconducting flux qubit capable of detecting individual photons propagating through a transmission line with a refresh rate of $\sim 400$~ns, a narrow bandwidth, a well-characterized efficiency of $\sim 0.66 \pm 0.06$, and a low dark count rate of $\sim0.01$. If such a detector with bandwidth $B$ is attached at the end of the transmission line in Fig.~\ref{fig1} and is designed to count propagating microwave photons with a central frequency $\W$, Eq.~\eqref{bbo} gives the expected rate of photons arriving at the detector as
\beq
\label{nomega}
N(\W)=n_B(\W,T_t)B+ \frac{E_\F}{\hbar\W} \frac{\S(\W,\mu)}{\hbar^2},
\eeq
where 
\begin{multline}
\label{bbw}
\S(\W,\mu)\equiv[1+n_B(\W,T_t)]dS(\W,\mu,T)\\
-n_B(\W,T_t)dS(-\W,\mu,T)\ .
\end{multline}
Here, we have assumed a high-quality resonator obeying $\ka\lesssim B\ll\W$. The first term in $N(\W)$ is the number of background thermal photons while the second term represents photon emission and absorption during interchain tunneling. 
In the limit of low transmission line/detector temperatures $k_BT_t\ll\hbar\W$, we have $n_B(\W,T_t)\ll 1$ and thus obtain 
\beq
\label{N}
N(\W)\approx\frac{E_\F}{\hbar\W}\frac{dS(\W,\mu,T)}{\hbar^2}\ .
\eeq
This is a central result of this work, i.e., the output photon flux gives a direct measurement of the ac spin current noise generated at the quantum magnet junction. 

\begin{figure}[t]
\includegraphics[width=0.8\linewidth]{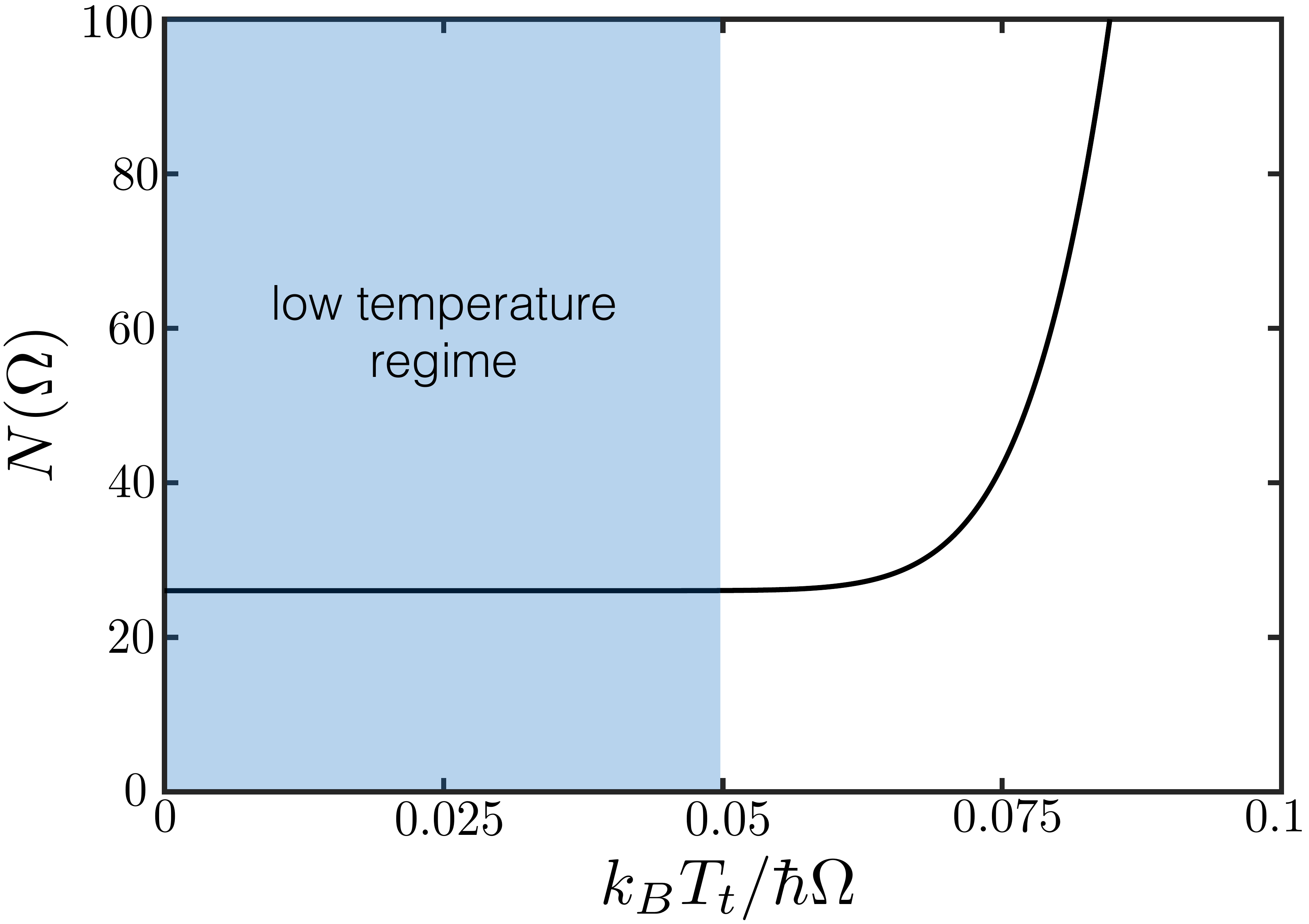}
\caption{A plot of $N(\W)$ as a function of transmission line temperature $T_t$. Here, we use $\W/2 \pi=1$~GHz and spin temperature $T = 4$~K to remain in the dc spin transport regime, and spin bias $\mu = 0.03J$. The plateau in the limit of small $T_t$ (shaded in blue) is the dc spin current noise. $N(\W)$ exhibits the same qualitative behavior for essentially all spin biases $\mu$, thus allowing one to quantify the noise using this extraction method for various values of $\mu$.}
\label{fig2}
\end{figure}

If the spin temperature is held higher than the resonance frequency, i.e., $\hbar\W\ll k_BT$, the output photon flux directly measures the dc component of the spin current noise, i.e., 
\beq
N\approx \frac{E_\F}{\hbar\W} \frac{dS(\W=0,\mu,T)}{\hbar^2}.
\eeq
Since the spin current $I$ across the junction can be measured directly using inverse spin Hall effect at the right metallic reservoir (see Fig.~\ref{fig1}), the proposed system allows one to extract a quantity directly proportional to the dc spin Fano factor, defined as the noise normalized by the current
\beq
F= \frac{dS(\W=0,\mu,T)}{\hbar I},
\eeq
recently studied in Refs.~\onlinecite{kamraPRL16,matsuoPRL18,aftergoodPRB18}.

Figure~\ref{fig2} shows a plot of photon flux $N(\W)$, Eq.~\eqref{nomega}, as a function of the detector temperature $T_t$. It exhibits the same qualitative behavior for essentially all spin biases $\mu$, thus allowing one to quantify the noise using this proposed extraction method for various values of $\mu$. The plot in the figure is generated for a spin tunnel junction with coupling strength $\xi_c\sim0.01J$ and spin bias $\mu=0.03J$, intrachain exchange scale of $J/k_B\sim10^3~\mathrm{K}$~\cite{hirobeNATP16}, spin temperature of $T=4$~K , and an inductor loop of inductance $L\sim1$~nH with dimensions $x_2-x_1=w=10~\mathrm{\mu m}$ placed $\de \sim1$~nm from spin chain 2. Under these conditions, we estimate $E_\F \sim 10^{-11}~\mathrm{K} \sim 2~\mathrm{Hz}$. For these parameters and at resonance frequency $\W/2\pi=1$~GHz~\footnote{Here, we assumed a cut off length in the Luttinger theory of $\eta/a\approx10$}, we expect the chain-to-cavity interaction to produce approximately 26 photons per second in the low temperature regime (shaded in blue), where the photon flux converges to the quantity in Eq.~\eqref{N}. We believe the detection of this number of output photons is within the reported capabilities of a single microwave photon detector impedance matched to a transmission line~\cite{inomataNATC16}. 


\section{Conclusion} 

We have shown that by placing a high-quality microwave resonator circuit in close proximity to a pair of exchange-coupled spin chains it is possible to directly measure spin transport quantities at a junction between the two spin systems. When the electromagnetic environment interacts weakly with the spin system, the setup allows measurement of both finite frequency and dc spin current fluctuations by examining the total number of output photons produced by interactions with the environment. This work opens doors to the possibilities of exploring the photodetection statistics of radiation produced at a junction between two quantum magnets and the generation of antibunched photons by exploiting the similarity between the current spin subsystem and a tunnel junction between two quantum conductors~\cite{beenakkerPRL01,*beenakkerPRL04}. Our theory can also be extended to describe spin transport between tunnel-coupled gapped spin systems (e.g., quantum Ising chains) that may mimic the dynamical Coulomb blockade physics of superconducting Josephson junctions coupled to electromagnetic environments~\cite{ingoldCHA92}. A final intriguing possibility would be to recast the detection methodology proposed here in terms of a generalized full counting statistics formalism as expressed in, e.g., Ref.~\onlinecite{nazarovEPJB03}.

{\it Acknowledgments.}~The authors thank J. Yuan, and M. Dartiailh for valuable discussions. This research was supported by Research Foundation CUNY Project \#90922-07~10.

\end{document}